\documentclass[%
reprint,
amsmath,
amssymb,
aps,
prx,
]{revtex4-2}

\usepackage{graphicx}
\usepackage{dcolumn}
\usepackage{bm,bbm}
\usepackage[colorlinks, citecolor=black, linkcolor=black, urlcolor=blue]{hyperref}
\usepackage{physics}
\usepackage{siunitx}
\def \eps{\varepsilon_c}
\def \Kappa{\mathcal{K}}
\usepackage{svg}

\usepackage{lipsum}

\begin{document}

\title{Challenging the Quantum Advantage Frontier with Large-Scale Classical Simulations of Annealing Dynamics}

\author{Linda Mauron}
\affiliation{
Institute of Physics, École Polytechnique Fédérale de Lausanne (EPFL), CH-1015 Lausanne, Switzerland
}
\affiliation{
Center for Quantum Science and Engineering, \'{E}cole Polytechnique F\'{e}d\'{e}rale de Lausanne (EPFL), CH-1015 Lausanne, Switzerland
}

\author{Giuseppe Carleo}
\email{giuseppe.carleo@epfl.ch}
\affiliation{
Institute of Physics, École Polytechnique Fédérale de Lausanne (EPFL), CH-1015 Lausanne, Switzerland
}
\affiliation{
Center for Quantum Science and Engineering, \'{E}cole Polytechnique F\'{e}d\'{e}rale de Lausanne (EPFL), CH-1015 Lausanne, Switzerland
}

\date{\today}

\begin{abstract}
    Recent demonstrations of D-Wave's annealing-based quantum simulators have established new benchmarks for quantum computational advantage [\href{https://doi.org/10.48550/arXiv.2403.00910}{arXiv:2403.00910}]. However, the precise location of the classical-quantum computational frontier remains an open question, as classical simulation strategies continue to evolve. Here, we demonstrate that time-dependent variational Monte Carlo (t-VMC) with a physically motivated Jastrow-Feenberg wave function can efficiently simulate the quantum annealing of spin glasses up to system sizes previously thought to be intractable. Our approach achieves accuracy comparable to that of quantum processing units while requiring only polynomially scaling computational resources, in stark contrast to entangled-limited tensor network methods that scale exponentially. For systems up to $128$ spins on a three-dimensional diamond lattice, we maintain correlation errors below $7\%$, which match or exceed the precision of existing quantum hardware. Rigorous assessments of residual energies and time-dependent variational principle errors establish clear performance benchmarks for classical simulations. These findings substantially shift the quantum advantage frontier and underscore that classical variational techniques, which are not fundamentally constrained by entanglement growth, remain competitive at larger system sizes than previously anticipated.
\end{abstract}

\maketitle

\section{Introduction}

More than $40$ years after R. Feynman's introduction of the concept of quantum computers \cite{feynman_simulating_1982, preskill_quantum_2023}, quantum processing units (QPUs) have become the focus of active research, with multiple platforms now developed \cite{king_observation_2018, ebadi_quantum_2021, scholl_quantum_2021, bravyi_future_2022, cong_hardwareefficient_2022, moses_racetrack_2023, maring_versatile_2024}. As their capacities have significantly increased, demonstrating the \textit{advantage} \footnote{In this work we refrain from using the expression \textit{quantum supremacy}, otherwise used in several published work.} of quantum simulators over classical ones has become a central objective in the field. These demonstrations have primarily followed two distinct paths: random quantum circuit sampling and quantum simulation.

Random circuit sampling experiments, pioneered by Google's Sycamore processor \cite{arute_quantum_2019}, have shown that specific quantum circuits can be executed on quantum hardware faster than their classical counterparts. Subsequent demonstrations \cite{zhong_quantum_2020, wu_strong_2021, zhu_quantum_2022} have further reinforced these claims, though debates about the classical simulation frontier continue, sparkled by the development of classical simulations techniques that match or even outperform current experiments \cite{pan_solving_2022}. In parallel, quantum simulation experiments have demonstrated the ability to prepare and probe complex many-body states \cite{kim_evidence_2023}, with recent work showing efficient preparation of multi-dimensional spin glasses up to unprecedented system sizes \cite{king_computational_2024}.

Notably, most quantum advantage claims rely on comparisons with either exact classical simulation methods or tensor network approaches. Exact methods, while providing a reliable baseline, inherently scale exponentially with system size. 
On the other hand, traditional tensor network methods, particularly those based on matrix product states (MPS) Ansätze, are particularly prone for the simulation of quantum circuits, since local operators can be directly absorbed into the parameters of the product-of-tensors wave function. However, they face fundamental limitations due to their limited ability to capture highly entangled states \cite{vidal_efficient_2003, eisert_entanglement_2013, eisert_quantum_2015}. Specifically, the popular MPS Ansatz suffers from an exponentially increasing demand for computational resources due to the area law scaling of entanglement entropy \cite{hastings_area_2007, schuch_entropy_2008, arad_area_2013}.

However, these comparisons leave room for alternative classical approaches that might circumvent both the exponential scaling of exact methods and the entanglement limitations of tensor networks. 
In this context, variational Monte Carlo, which imposes virtually no restriction on the structure of the variational Ansatz, emerges as a promising candidate for ground state search (VMC) and unitary time evolution (t-VMC) \cite{carleo_localization_2012, schmitt_dynamical_2015, ido_timedependent_2015, carleo_unitary_2017, schmitt_quantum_2020, mauron_predicting_2024, carleo_simulating_2024}. The versatility of the Ansatz, ranging from simple mean-field models to deep neural quantum states \cite{carleo_solving_2017}, allows it to be tailored to any system or design feature. Notably, correlations can be explicitly encoded within the structure of the Ansatz. Moreover, even for complex structures, the use of advanced numerical tools, such as automatic differentiation, ensures efficient optimization. 
Additionally, the use of Monte Carlo Markov Chain (MCMC) sampling ensures that the efficiency of the method persists as system size increases.

In this work, we employ t-VMC to simulate the quantum annealing of a spin glass state. Using a variational Ansatz based on the Jastrow-Feenberg expansion, we demonstrate that our method achieves comparable accuracy to QPUs with time resources only scaling polynomially in the system size. 
The structure of our work is as follows. We start by defining the quantum annealing protocol and physical system in Sec. \ref{sec:protocol}. Subsequently, Sec. \ref{sec:method} introduces our numerical method, namely the t-VMC prescription as well as the variational wave function. The accuracy and efficiency of our method are presented in Sec. \ref{sec:results}, and finally discussed in Sec. \ref{sec:conclusion}. 

\section{\label{sec:protocol} Quantum annealing }

In many complex optimization problems, the presence of multiple local minima makes accessing the global minimum a tremendous challenge. Specifically, the loss or energy landscape often makes optimization impracticable through conventional methods.
To address such challenges, a quantum annealing process \cite{santoro_theory_2002, das_colloquium_2008, heim_quantum_2015} offers a practical solution. This method takes advantage of quantum fluctuations by initializing the quantum state as the ground state of an easy driving Hamiltonian $\hat{\mathcal{H}}_D = -\sum_i \hat\sigma_i^x$, where $\hat{\sigma}_i^x$ is the $x$-axis Pauli operator acting on site $i$. The system then undergoes time evolution under the time-dependent Hamiltonian 
\begin{equation}
    \mathcal{\hat{H}}(t) = \Gamma(t)\hat{\mathcal{H}}_D + \Kappa(t)\hat{\mathcal{H}}_T\, \text{.}
\label{eq:H}
\end{equation}
This evolution will result in the ground state of the target Hamiltonian $\hat{\mathcal{H}}_T$ if $\Gamma(0)\gg\Kappa(0)$ and $\Gamma(T)\ll\Kappa(T)$ for a sufficiently long annealing time $T$ \cite{kato_adiabatic_1950, bapst_quantum_2013}.

Among systems with complex energy landscapes, spin glasses represent a canonical example \cite{edwards_theory_1975}. 
They are characterized by neighboring pairs of spins that interact via random couplings $J_{ij}$. The resulting Edwards-Anderson Hamiltonian 
\begin{equation}
    \hat{\mathcal{H}}_T = \sum_{\langle ij\rangle} J_{ij} \hat{\sigma}_i^z\hat{\sigma}_j^z\, \text{, } J_{ij}\sim U(-1,1)\,\text{, }
\label{eq:HT}
\end{equation}
where $\langle i j \rangle$ denotes all the edges of the lattice, serves as a paradigmatic case of disordered systems. The strong frustration arising from the random couplings, which follow the uniform distribution $U(-1,1)$, results in a rugged and highly degenerate energy landscape.

Given these characteristics, the preparation of a spin glass ground state represents a compelling application for quantum annealing. Building on this approach, King \textit{et al.} \cite{king_computational_2024} recently conducted a comprehensive study of spin glass characteristics across various lattice geometries. Their experimental implementation on D-Wave's quantum processors systematically investigated fundamental properties of these systems, enabling them to achieve unprecedented system sizes and statistical sampling. They thus concluded on the advantage of quantum simulations over classical ones based on MPS simulations.

\section{\label{sec:method} Variational approach}

As previously introduced by Ref. \cite{carleo_simulating_2024}, to realize a classical simulation of the annealing protocol, we approximate the wave function using a variational Ansatz 
\begin{equation}
    \ket{\psi(t)} = \sum_{\{\bm{\sigma}\}} \psi_{\bm{\theta}(t)}(\bm{\sigma}) \ket{\bm{\sigma}}\,\text{,}
\end{equation}
where the states $\{\bm{\sigma}\} = \{ \uparrow, \downarrow\}^{\otimes N}$ span the whole Hilbert space of $N$ spins and $\bm{\theta}(t)$ are the time-dependent variational parameters. 
Using Monte Carlo Markov Chains, it is possible to sample the Born distribution associated to the variational wave function $p(\bm{\sigma};t) = |\psi_{\bm{\theta}(t)}(\bm{\sigma})|^2/||\psi_{\bm{\theta}(t)}||^2$. Moreover, using Monte Carlo integration, an operator $\hat{O}$ can be efficiently evaluated as $\ev*{\hat{O}} = \mathbbm{E}_{\bm{\sigma}\sim p} \left [ O_\text{loc}(\bm{\sigma})\right ]$, which is an average of the so-called \textit{local} operator $O_\text{loc}(\bm{\sigma}) = \sum_{\bm{\sigma}^\prime} \mel{\bm{\sigma}}{\hat{O}}{\bm{\sigma}^\prime} \frac{\psi_{\boldsymbol{\theta}}(\bm{\sigma}^\prime)}{\psi_{\boldsymbol{\theta}}(\bm{\sigma})}$. Here, the sum runs over the configurations $\bm{\sigma}'$ for which the matrix elements $\mel{\bm{\sigma}}{\hat{O}}{\bm{\sigma}^\prime}$ are non-zero. This local estimator can be efficiently computed for most cases of physical interest, including arbitrary diagonal operators, and for any $k$-local operator.

In this variational framework, the time evolution of the wave function is governed by the time dependence of its parameters. The time-dependent variational principle (TDVP) determines the optimal time dependence by minimizing the Fubini-Study distance between the evolved state and a state with updated parameters $\mathcal{D}\left( e^{-i\hat{\mathcal{H}}(t)dt} \ket{\psi_{\bm{\theta}(t)}}, \ket{\psi_{\bm{\theta}(t) + \dot{\bm{\theta}}(t)dt}} \right)$ at each time step. 
Relying on the same equation of motion
\begin{equation}
    \sum_{k^\prime} S_{kk^\prime} \dot{\theta}_{k^\prime}(t) = -i F_{k} \,\text{,}
\label{eq:t-VMC}
\end{equation}
the time-dependent variational Monte Carlo (t-VMC) prescription \cite{carleo_localization_2012, yuan_theory_2019} evaluates both the quantum geometric tensor $S_{kk^\prime} = \langle\hat{D}_k^* \hat{D}_{k^\prime}\rangle - \langle\hat{D}_k^*\rangle \langle\hat{D}_{k^\prime}\rangle$ and the forces $F_k = \langle\hat{D}_k^* \hat{\mathcal{H}}\rangle - \langle\hat{D}_k^*\rangle \langle\hat{\mathcal{H}}\rangle$ using Monte Carlo integration. Here, the local derivatives operator is defined as $\hat{D}_i\ket{\psi_{\bm{\theta}}} = \ket{\partial_{\theta_i} \log\psi_{\bm{\theta}}}$.

The accuracy of this method can be estimated through the second order approximation of the Fubini-Study distance
\begin{equation}
    r^2(t) = 1+\frac{\dot{\bm{\theta}}^{\dagger}(t)(\bm{S} \dot{\bm{\theta}}(t) + i \bm{F}) - i \bm{F}^{\dagger} \dot{\bm{\theta}}(t)}{(\delta E)^2}\,\text{,}
\label{eq:r2}
\end{equation}
which provides an estimate of the TDVP error \cite{schmitt_quantum_2020, hofmann_role_2022}. Here, $(\delta E)^2$ is the variance of the energy estimate. The total error of the t-VMC scheme is then obtained by integrating this quantity over time $R^2 = \int_0^T dt ~ r^2(t)$. Details on the evaluation of this quantity can be found in App. \ref{app:R2}.

\subsection{Time-Dependent Jastrow States}
\begin{figure*}[ht]
    \includegraphics[width=\linewidth]{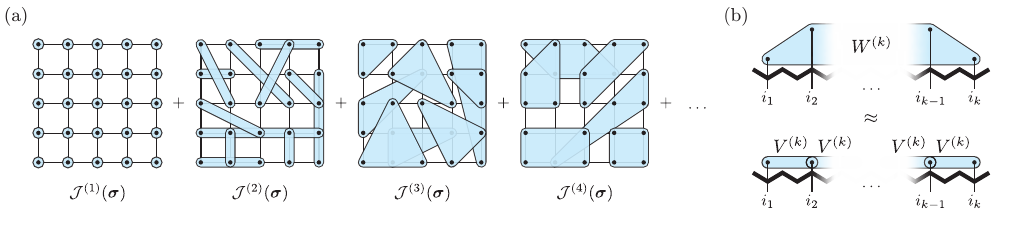}
    \caption{Representation of the variational Ansatz. 
    (a) Representation of the Jastrow-Feenberg expansion defined in EQ.\eqref{eq:expansion} up to $4$-th order. The blue boxes stand for examples of $k$-body tuples on a square lattice, where the spins considered are represented as black circles. For each Jastrow correlator of order $k$, all $k$-body tuples are used in the summation \eqref{eq:jastrow}. 
    (b) Illustration of the factorized $k$-body Jastrow parameter as defined in Eq. \eqref{eq:W4}. The indices $i_2$ up to $i_{k-1}$ are not contracted before the input of the state $\bm{\sigma}$. The thick zigzags stand for the symmetric property of the tensors, following the notation introduced by Penrose \cite{penrose_applications_1971}. }
    \label{fig:ansatz}
\end{figure*}

Within the t-VMC framework, choosing an appropriate Ansatz is crucial for accurately representing the quantum state. Given the spin-spin interactions inherent to the Hamiltonian \eqref{eq:H}, the Jastrow-Feenberg expansion emerges as a promising candidate due to its natural ability to encode correlations \cite{jastrow_manybody_1955, cevolani_protected_2015}. This Ansatz is formulated using $k$-body correlators of the form
\begin{equation}
    \mathcal{J}^{(k)}(\bm{\sigma};t) = \sum_{i_1<\dots<i_k} W_{i_1\dots i_k}^{(k)}(t) \sigma_{i_1} \dots \sigma_{i_k}\,\text{,}
\label{eq:jastrow}
\end{equation}
which entangle spins together through the complex variational parameter $\bm{W}^{(k)}(t)$. This Ansatz is a robust choice for the t-VMC algorithm, given the simplicity of the local derivatives calculations. Moreover, any expansion of order $p$ 
\begin{equation}
    \psi_{\bm{\theta}(t)}^{(p)}(\bm{\sigma}) = \exp\left[  \sum_{k=1}^p\mathcal{J}^{(k)}(\bm{\sigma};t) \right] \, \text{,}
\label{eq:expansion}
\end{equation}
can be systematically improved by including higher order terms. 
While an expansion of order $p=N$ is exact, in practice, a lower-order expansion, as illustrated in Fig. \ref{fig:ansatz}, provides sufficient accuracy for most applications. In particular, given that the two-body Jastrow is capable of exactly representing the ground state of both the driving ($W^{(2)}_{ij} = 0$) and the target Hamiltonian ($W^{(2)}_{ij} \propto J_{ij}$), we anticipate that a second-order expansion already constitutes a good quantitative approximation for the time-dependent state. 

Yet, in order to improve the representational power during time evolution, we wish to include higher-body correlators in the Ansatz as well. 
However, the number of parameters in higher-body Jastrow tensors scales as $\mathcal{O}(N^k)$, making it impractical for large numbers of spins. We thus seek a lower-rank version, featuring a small number of parameters, while retaining the accuracy of the full $k$-body term. Specifically, we consider the rank-$2$ approximation
\begin{equation}
    W_{i_1 \dots i_k}^{(k)} = V_{i_1 i_2}^{(k)} \cdot V_{i_2 i_3}^{(k)} \dotsb V_{i_{k-1} i_k}^{(k)}\,\text{,}
\label{eq:W4}
\end{equation}
which provides a compact and efficiently evaluated representation of $k$-body correlators, while maintaining the symmetry of the original Jastrow form. This property directly results from the form of the correlator provided in Eq. \ref{eq:jastrow}, which only sums terms in one single symmetry sector ($i_1 < \dots < i_k$).
With this choice, evaluating the $k$-body Jastrow correlator requires an effort comparable to the two-body term, with computational scaling $\mathcal{O}( N^2 )$.

\begin{figure}[ht]
\centering
    \includegraphics[width=\linewidth]{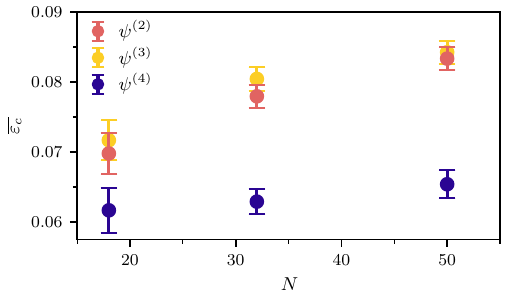}
    \caption{ Averaged correlation error for different Ans\"atze. Each color corresponds to a different order of the Jastrow-Feenberg expansion \eqref{eq:expansion}. 
    For all Ans\"atze, the data obtained is compared to QPU data of Ref. \cite{king_computational_2024}. Error bars are estimated as the root-mean-squared error over $20$ the random spin glass realizations. }
\label{fig:comparison}
\end{figure}

In order to choose the order of the expansion suitable for our problem, we investigate the accuracy of the final state for increasing expansion order. 
Given that most order parameters of spin glasses are based on spin-spin correlations $c_{ij} = \ev{\hat{\sigma}_i^z \hat{\sigma}_j^z}$, we use the correlation error 
\begin{equation}
    \eps = \left(\frac{\sum\limits_{i<j} (c_{ij}-\widetilde{c}_{ij})^2}{\sum\limits_{i<j} \widetilde{c}_{ij}^{~2}} \right)^{1/2}
\label{eq:epsilon}
\end{equation}
with regards to some reference data $\widetilde{c}_{ij}$ as a measure for precision on the final state. 
Due to the random nature of this model, different realizations of the spin glass Hamiltonian can present drastically different behavior. Therefore, we study the average of multiple coupling realizations, a quantity that we denote $\overline{\eps}$.

In Fig. \ref{fig:comparison}, we thus present the correlation error for increasingly large systems using the second-order Jastrow expansion. Despite the simplicity of this approximation, the error remains in the order of $10^{-2}$ for the systems analyzed, demonstrating high precision. However, because of the presence of intermediate time-evolved states that may not be captured accurately by a two-body Jastrow Ansatz alone, some imprecision persists, suggesting that a higher-order expansion might be necessary.

Interestingly, we have found in Fig. \ref{fig:comparison} that the inclusion of three-body correlators does not improve precision. This is likely due to the fact that excitations generated during the dynamics are not dominantly three-body. Thus, considering the increased complexity due to the higher number of parameters, the third-order expansion performs marginally worse than the second-order one. This leads us to exclude three-body correlations from our variational model and instead concentrate on four-body correlations.

Using the fourth-order form, Fig. \ref{fig:comparison} shows a direct improvement over the two-body Jastrow. Even for small systems, we observe a significant reduction in error with the higher-order Jastrow Ansatz. Additionally, this approach mitigates the growth of errors with increasing system size, ensuring consistent improvement in precision. Thus, we adopt this four-body Jastrow form as the main variational wave function to be used in the numerical experiments performed in the next sections.

\section{\label{sec:results} Results}
In this section, we present the numerical results of the quantum annealing presented in Sec. \ref{sec:protocol} using the t-VMC method. 
Given that higher dimensionality and connectivity of the lattice tend to increase frustration and thus raise the complexity of numerical simulation, we decide to focus our study to the highest-dimensional lattice available on Ref. \cite{king_computational_2024}'s QPUs. Further restricting to dimerless lattices \footnote{Given that the construction of the dimerized lattices relies on strong ferromagnetic couplings $J_{\text{dimer}}=-2$, two different energy scales are present in these systems. This fact can impact the t-VMC integration scheme, which is why we leave the study of such systems to future work. } leaves us with the diamond lattice, which is presented in more details in App. \ref{app:lattice}.

\begin{figure}[ht]
    \centering
    \includegraphics[width=\linewidth]{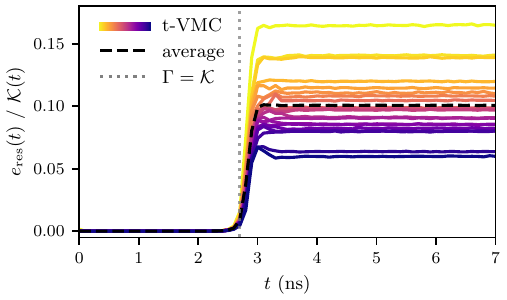}
    \caption{ Residual energy per spin $e_\text{res} = E_\text{res}/N$, renormalized by the value of the frequency $\Kappa$ at all times. The data is obtained on the $N=18$ diamond lattice, where the ground state energy $E_0$ can be computed exactly, for an annealing process of $T=\SI{7}{\ns}$. Each color represents a different realization of the spin glass weights $J_{ij}$ and their average is depicted by the black dashed line. The point in time where $\Gamma(t) = \Kappa(t)$ is shown as a dotted line. }
    \label{fig:residual}
\end{figure}

For an infinitely slow protocol, the annealing process is perfectly adiabatic, implying that the system's energy coincides with the instantaneous ground state energy $E_0(t)$. In small systems, the amount of nonadiabaticity can be monitored by the behavior of the residual energy $E_\text{res}(t) = \langle\hat{\mathcal{H}}(t)\rangle - E_0(t)$, which is nonvanishing in the presence of nonadiabatic passages. During the initial phase of the evolution, the zero-valued residual energy, presented in Fig. \ref{fig:residual}, suggests near-perfect adiabaticity However, at $t\approx\SI{2.75}{\ns}$, the state crosses the critical point and diverges from the instantaneous ground state, thus yielding a finite residual energy. Subsequently, the  evolution stabilizes and stops diverging, thus evolving the state without further exciting it. During the whole process, the residual energy remains low, indicating that the state is still close to the ground state.

\subsection{Correlation error}
\begin{figure*}
    \centering
    \includegraphics{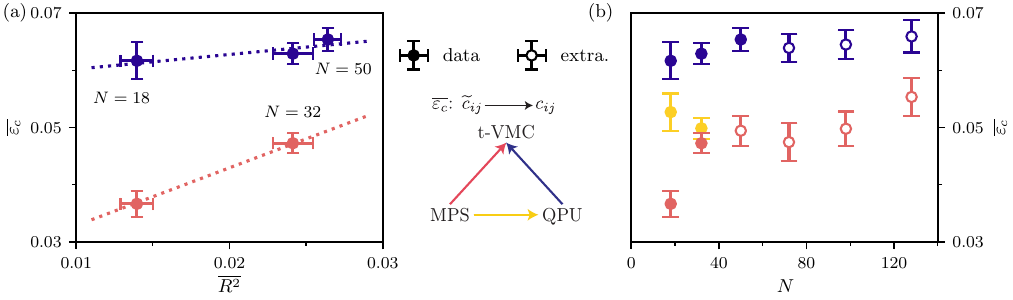}
    \caption{Averaged correlation error. Each color indicates the data in comparison, as illustrated in the center legend, where an arrow of the corresponding color links the reference values (base of arrow) to some tested data (tip). 
    Filled points (data) represent the systems for which reference data was available and the correlation error $\eps$ could be exactly calculated, as the empty ones (extra.) are issued from linear extrapolation. 
    (a) Linear relation between the average TDVP error estimate and the correlation error. (b) Resulting correlation error for growing system sizes. Yellow points are the results presented in Ref. \cite{king_computational_2024}. }
    \label{fig:epsilon}
\end{figure*}

To assess the accuracy of our method, we consider the QPU results from Ref. \cite{king_computational_2024} as reference values for small systems ($N\leq50$). As Fig. \ref{fig:epsilon}(a) demonstrates, the correlation error averaged over $20$ spin glass realizations $\overline{\eps}$ is directly proportional to the average TDVP error $\overline{ R^2 }$. Given that the TDVP error is accessible for any t-VMC simulation, this relation allows for the direct extrapolation of $\overline{\eps}$, even in the absence of reference values.

Using this approach, we show in Fig. \ref{fig:epsilon}(b) that after evolution with t-VMC, the average error remains strictly below $7\%$ for systems up to $N=128$ spins. Although a slight increase with system size is observed for smaller systems, the average error approaches a stable regime where it is approximately constant, thus the error per site decreases significantly.
Moreover, we find errors of the same order of magnitude as those reported in Ref. \cite{king_computational_2024} (yellow in Fig. \ref{fig:epsilon}), confirming the sufficient accuracy of our method.

Furthermore, for a comprehensive analysis, we compare our approach to some other numerical method, namely tensor network simulations based on MPS Ans\"atze, as it is one of the leading approaches for time-dependent quantum simulations \cite{schmitt_quantum_2022}. 
Using the two data points provided in Ref. \cite{king_computational_2024}, we conducted the same analysis as for the QPU data \footnote{The MPS calculations in Ref. \cite{king_computational_2024} were performed for only two lattice sizes ($N=18$ and $N=32$), and thus, the previously observed linear relation cannot be proved. Nevertheless, based on the observations from the QPU data, we present the extrapolated results using MPS data, while cautioning the reader that the linear regression is based on only two data points.}. 
In this case, the qualitative behavior is similar to the comparison with QPU data, presenting, however, a lower correlation error for all system sizes. The better agreement between our simulations and well converged MPS simulations is likely explained by the imperfections present in the QPU, leading to a non-unitary dynamics that correlates worse with our purely unitary-dynamics simulations. 

\subsection{Computational cost}
\begin{figure}[h!]
    \centering
    \includegraphics{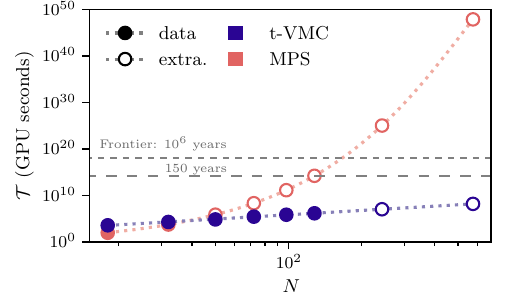}
    \caption{ Total time cost (in GPU seconds) of a time evolution simulation. The data for the t-VMC simulations is measured for smaller systems ($N\leq128$) and follows a polynomial law, which allows for extrapolation to larger systems. In the case of MPS, the time of simulation is calculated from the reported bond dimension $\chi$ as $\mathcal{T} \propto T N \chi^3$. The extrapolated data stands for bond dimensions extrapolated through the area law $\ln \chi \propto N^{2/3}$. The horizontal dashed lines represent the total computation time available on the Frontier clusters during $150$ years and $1$ million years respectively \cite{bethea_ornl_2022}. }
    \label{fig:cost}
\end{figure}

In addition to accuracy, the cornerstone of an efficient simulation resides in the computational resources required. 
We show in Fig. \ref{fig:cost} the measured time $\mathcal{T}$ needed to perform the t-VMC simulations presented in this work. In particular, we find that the total time cost for simulating the annealing protocol exhibits only polynomial growth $\mathcal{T}\sim T N^3$, which allows us to simulate lattices of up to $N=128$ spins in only a few days using a few (up to $4$) GPUs. 

In contrast, the bond dimension of an MPS grows as $\ln(\chi) \sim N^{2/3}$ due to area-law, leading to a time cost of $\mathcal{T} \sim T N \exp(3N^{2/3})$. This almost-exponential growth leads to severe restrictions on the system sizes available to TNs simulations. 
For instance, we estimate that an MPS-simulation of our largest system ($N=128$) would already require $150$ years of calculations using the full GPU capacity of \textit{Frontier}'s supercomputers \cite{bethea_ornl_2022}. 
In contrast, under the same conditions, simulating the largest system studied by the QPU in Ref. \cite{king_computational_2024} ($N=576$) would require a few hours using our t-VMC scheme, against $10^{35}$ years for MPS.

\section{\label{sec:conclusion} Discussion and Conclusions}

In this work, we propose time-dependent variational Monte Carlo as a method to simulate quantum annealing of a spin glass state. In doing so, we are able to accurately and efficiently simulate faster-than-adiabatic dynamics of a quantum simulator. Unlike traditional tensor network techniques, our correlated Ansatz is not limited by the growth of entanglement and, at fixed expansion order, only requires polynomially increasing computational resources for simulation. 

Although our approach gives an exponential speed-up compared to MPS methods, it is worth mentioning that quantum processing units still provide a computational advantage in terms of total running time, provided one is willing to sacrifice accuracy when compared to a purely unitary time evolution. However, our method has the advantage of being very flexible, both in the type of Hamiltonian taken into consideration and the choice of the lattice. Moreover, our Ansatz being based on the Jastrow-Feenberg expansion \eqref{eq:jastrow}, its accuracy can systematically be improved by increasing the order of the Jastrow representations, at the cost, however, of the number of parameters and thus computational resources. 

Most importantly, our findings call for a significant shift in the computational "advantage" frontier identified in Ref. \cite{king_computational_2024}. 
These findings challenge prevailing narratives on quantum advantage grounded solely in entanglement-based arguments and entangled-limited classical simulations. By demonstrating that variational methods like Jastrow-Feenberg expansions can feasibly capture such entanglement at polynomial cost, we significantly shift the boundaries of what was deemed computationally intractable for adiabatic quantum simulation. We anticipate further advances through alternative, non–entanglement-limited classical ans\"atze—among them neural quantum states—which promise new avenues for exploring the efficiency and accuracy of adiabatic and faster-than-adiabatic quantum state preparation.

\begin{acknowledgments}
The simulations in this work were carried out using NetKet \cite{carleo_netket_2019, vicentini_netket_2022}, which is based on Jax \cite{bradbury_jax_2018} and MPI4Jax \cite{hafner_mpi4jax_2021}. 
This research was supported by SEFRI through Grant No. MB22.00051 (NEQS - Neural Quantum Simulation).

\textit{Note: During the preparation of this manuscript, we became aware of a related work by Tindall }et al.\textit{ simulating the same protocol using tensor networks \cite{tindall_dynamics_2025}. }
\end{acknowledgments}

\appendix
\numberwithin{table}{section}
\numberwithin{figure}{section}
\section{\label{app:R2} TDVP Error estimate}

Given that all observables are sensitive to stochastic noise in the t-VMC scheme, the equation of motion, which depends on $S_{kk^\prime} = \mathbbm{E}[D^*_k ( D_{k^\prime} - \mathbbm{E}[D_{k^\prime}] )]$ and $F_k = \mathbbm{E}[D^*_k ( E_\text{loc} - \mathbbm{E}[E_\text{loc}] )]$, can be impacted by sampling. 
Therefore, if the MCMC sampling is not performed perfectly, the equation of motion will depend on the specific samples and, thus, the parameters $\bm{\theta}(t)$ will be biased. However, in such a case, the TDVP error \eqref{eq:r2} will be low, since it depends on exactly the same estimates.

Thus, a noise-sensitive measure is needed to unambiguously characterize the quality of the time-evolution scheme, including sampling. This can be achieved by considering the \textit{validation error} $r^2(\bm{\dot{\theta}}^{(1)}, \bm{S}^{(2)}, \bm{F}^{(2)}, \delta E^{(2)})$, where the superscripts $(i)$ indicate which set of samples is used for the estimation of each quantity. 
Relying on two independent sets of samples, it allows one to characterize the error occurring from over-fitting to noise, supplementary to other linear algebra errors. 
In general, the validation error will measure higher errors than the usual estimator, obtained with a single set of samples. The same results are obtained for a perfectly well-behaved sample-set.

\section{\label{app:lattice} Diamond lattice}

\begin{figure}
    \centering
    \includegraphics[width=0.65\linewidth]{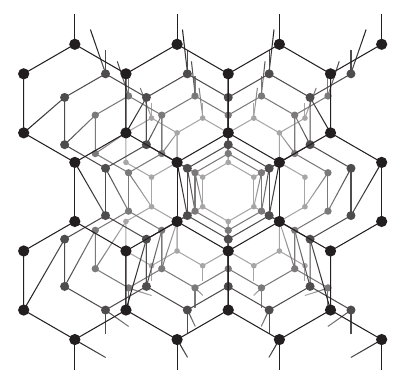}
    \caption{ Representation of the largest diamond lattice used during the simulations ($N=128$). The half links at the top and bottom of the figure represent the periodic boundary condition, only present in the vertical direction. Both other directions possess open boundaries. }
    \label{fig:lattice}
\end{figure}

In this work, we consider a diamond lattice as our system of spins. This three-dimensional lattice, depicted in Fig. \ref{fig:lattice}, possesses a periodic boundary in the vertical direction to reduce finite-size effects and otherwise open boundaries. 
The high connectivity of the lattice induces strong frustration, which represents a great challenge to simulate. 
In our simulations, we kept this periodic dimension to a fixed length and identically modified the two other sizes, ranging from $18$ spins to $128$.

\section{Monte Carlo sampling and Parallel Tempering}

\begin{figure*}[ht]
    \centering
    \includegraphics[width=\linewidth]{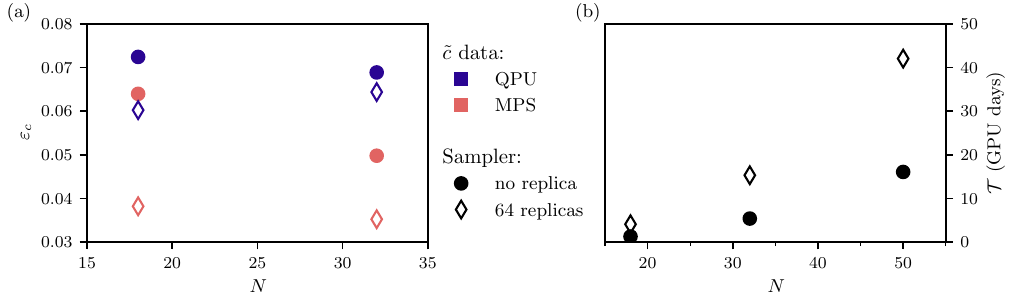}
    \caption{ Comparison between standard sampler (no replica) and parallel tempering ($64$ replicas) sampler using local-spin-flip transitions. The results are based on the simulation of a single realization of the random couplings in the Hamiltonian. For the comparison to be fair, the thermalization time was set to the same value for both sampling schemes. 
    (a) Correlation error with regards to QPU (blue) and MPS (red) data. 
    (b) Total time of evolution. The data for $N=50$ is extrapolated from a short-time simulation. }
    \label{fig:pt_sampler}
\end{figure*}

The efficiency of t-VMC relies significantly on the Monte Carlo Markov Chain (MCMC) approach, which involves the sampling of multiple chains generated by a random walk.
In each Markov chain, the probability of transitioning from a sample $\bm{\sigma}$ to a new one $\bm{\sigma}^\prime$ is given by 
\begin{equation*} 
    P(\bm{\sigma} \to \bm{\sigma}^\prime) = \min\left \{ 1, \frac{p(\bm{\sigma}^\prime)}{p(\bm{\sigma})} \right\}\,\text{,} 
\end{equation*}
where $p(\bm{\sigma}) = |\psi(\bm{\sigma})|^2$ is the Born probability distribution of the wave function \cite{becca_quantum_2017}.
Thus, for a transition rule that acts only on local degrees of freedom, such as the flipping of a single spin, only few transitions are accepted in the glassy phase, due to the peaked structure of the ground state, which consists of multiple disconnected spin configurations.

To address the challenge of glassy phases, \textit{parallel tempering} (PT) approaches, which are based on the effect of finite temperatures on the probability distribution of a state, have proven particularly helpful \cite{swendsen_replica_1986, hukushima_exchange_1996}.
In this framework, multiple \textit{replicas} of the system are simulated, each with a different (finite) temperature. Thus, similar to standard MCMC, a replica with inverse temperature $\beta$ will sample the modified probability distribution $p(\bm{\sigma}; \beta) = (|\psi(\bm{\sigma})|^2)^\beta$, which results in the transition probability
\begin{equation*} 
    P(\bm{\sigma} \to \bm{\sigma}^\prime) = \min\left \{ 1, \left(\frac{|\psi(\bm{\sigma}^\prime)|^2}{|\psi(\bm{\sigma})|^2}\right)^{\beta} \right\}\,\text{.} 
\end{equation*}
The main difference with the usual sampling scheme occurs at the next step, during which two replicas in state $(\beta,\bm{\sigma})$ exchange their temperatures with probability
\begin{equation*} 
    P(\beta_i \leftrightarrow \beta_j) = \min\left \{ 1, \left(\frac{|\psi(\bm{\sigma}_i)|^2}{|\psi(\bm{\sigma}_j)|^2}\right)^{\beta_i-\beta_j} \right\}\,\text{.} 
\end{equation*}
This exchange allows the Markov chains to explore the different domains of the glassy wave function, rather than remaining trapped in one.

Therefore, to assess the impact of sampling on the final state of the evolution, we compare, in Fig. \ref{fig:pt_sampler}(a), the correlation error obtained with each of these sampling methods. We observe that, regardless of the reference data, the parallel tempering approach obtains improved accuracy compared to the standard sampling mechanism. However, this improvement decreases already for a system of size $N=32$. Since we restricted this analysis to small systems, in order to compare with MPS data, we will not predict the impact on larger system sizes.

Nevertheless, we can draw conclusions regarding the efficiency of both schemes by comparing their computational costs.
While the parallel tempering method has proven to be particularly accurate for the study of spin glasses, its requirement for many ($64$) replicas increases the number of Markov chains to sample and, thus, the computational cost. As shown in Fig. \ref{fig:pt_sampler}(b), the resources required increase more rapidly in the case of PT sampling, even though the resulting accuracy does not further improve with system size.
Moreover, the replica-less simulations run in this analysis already exhibit an increased computational cost, compared to the simulations from the main text.
Specifically, since we observed that PT samplers required a larger number of thermalization steps, we decided to use the same thermalization for all simulations, for the sake of fairness. This increased the computational cost for the standard sampler, even though the accuracy was only slightly impacted.

We can thus conclude that the PT sampler provides greater accuracy than the standard Metropolis sampler. However, this improvement is moderate and does not drastically change the conclusions made in the main text concerning accuracy compared to MPS simulations. Furthermore, its increased complexity, especially in terms of simulation time, would likely make it impractical for simulating large systems. Consequently, we opted to sacrifice the slight improvement in accuracy for the faster solution, and chose to use the standard Metropolis sampler. However, it is worth noting that further improvements in quality could be achieved by modifying the sampling scheme in the t-VMC simulations.

\clearpage
\newpage
\bibliography{main}

\end{document}